\documentclass[a4paper,twocolumn,english,prb,article,endfloat]{revtex4}
\usepackage{array}
\usepackage{multirow}
\usepackage{amsmath}
\usepackage{graphicx}
\usepackage{bm}
\usepackage{hyperref}  
\usepackage{natbib}  

\begin{document}

\title{Untangling the role of oxide in Ga-assisted growth of GaAs nanowires on Si substrates}

\author{F. Matteini, G. Tutuncuoglu, D. R\"uffer, E. Alarcon-Llado}
\affiliation{Laboratoire des Mat\'eriaux Semiconducteurs, \'Ecole Polytechnique F\'ed\'erale
de Lausanne, 1015 Lausanne, Switzerland}
\author{A. Fontcuberta i Morral}
\affiliation{Laboratoire des Mat\'eriaux Semiconducteurs, \'Ecole Polytechnique F\'ed\'erale
de Lausanne, 1015 Lausanne, Switzerland}

\date{Received xx July 2013;}
\begin{abstract}
The influence of the oxide in Ga-assisted growth of GaAs nanowires on Si substrates is investigated.
Three different types of oxides with different structure and chemistry
are considered. We observe that the critical oxide thicknesses needed
for achieving nanowire growth depends on the nature of oxide and how
it is processed. Additionally, we find that different growth conditions
such as temperature and Ga rate are needed for successful nanowire
growth on different oxides. We generalize the results in terms of
the characteristics of the oxides such as surface roughness, stoichiometry
and thickness. These results constitute a step further towards the
integration of GaAs technology on the Si platform. 
\end{abstract}
\maketitle

\section{Introduction}

The interest for semiconducting nanowires has continuously increased
in the past decade because of their wide range of possibilities both
in applied and fundamental science.\cite{Czaban:2009qw,Anonymous:DMWvgapA,Kelzenberg:2010fa,Tian:2007kl,Qian:2008cv,Wallentin:cp,Heiss:2013jw,Krogstrup:mxvdUTp+,Mourik:2012je,Deng:2012gn,Schmidt:2006pn,Dey:2012ie}
Within the range of possible applications III/V semiconductors are among the
most promising materials. The small footprint of nanowires
enables the virtually defect-free integration of mismatched materials,
which would not be possible in the thin film form. \cite{Bessire:2011gq,Ohlsson:2002qK}
Moreover, III-V nanowires can be obtained on Si, providing a path for combining
the III-V and Si platforms. As nanowires start growing
generally in a single nucleation event followed by a layer-by-layer
mode, III-V nanowires grown on Silicon appear also free from anti-phase
boundaries otherwise often found in thin film counterparts.\cite{Dubrovskii:2008tu,Uccelli:2011tc,Fang:1990r[}
 One of the most successful ways of growing nanowires is the vapor-liquid-solid
method (VLS) in which a liquid droplet (denominated as catalyst) is
used for the gathering and preferential decomposition of growth precursors.\cite{Wagner:1964hh,Dubrovskii:2009sm} Upon supersaturation of this droplet, precipitation occurs at the interface with the substrate
in the form of nanowire. One of the most successful catalysts used
for VLS is gold. However, when heating a Si substrate with gold on top, the gold droplets incorporate a significant amount of Silicon by the formation of an eutectic. As a consequence, the growth of III-V nanowires on Si using gold is quite more challenging than on GaAs. Therefore, many groups working on the growth of III-V nanowires on Silicon have looked for alternative methods, including the selective area epitaxy and Ga-assisted growth of GaAs nanowires.\cite{Colombo:2008ci,Paek:2009kr,Sadowski:JSQlMZtc,Heiss:2008ff,Uccelli:2011tc,Jabeen:2008tZ,Breuer:2011gx,Anonymous:2012gc,Krogstrup:2013tl}
Since the first self-catalyzed growth of GaAs nanowires was achieved,
the nature of the oxide has been an important parameter in nanowire
growth.\cite{Krogstrup:2010ux,Plissard:2010qO,Plissard:YNqQ9GDR}
To date, few reports show successful growth without the presence of oxide on the substrate surface.\cite{Samsonenko:2011rN,Jabeen:2008tZ,Plissard:2010qO}
In all of these cases, there was a non-negligible time lapse between substrate preparation and loading in the ultra-high-vacuum environment. It is well established that Si surfaces naturally undergo oxidation even
at room temperature simply by exposing them to air. The same oxidation
process takes place also in the case of Hydrogen passivated surfaces.\cite{Neuwald:sk2aXoay,Burrows:1988hl,Soria:2012gm} As a consequence,
what is claimed as oxide-free surface might not have been so. One should also note that most of the works aiming at the understanding of the role of oxide
in the growth of GaAs nanowires by the Ga-assisted method were mostly performed on GaAs substrates.\cite{Anonymous:jlM9QbYX}
It was observed that the oxide thickness plays a role in achieving
nanowires with an epitaxial relation with the substrate or even achieving
growth at all. Interestingly, the reported ``critical\textquotedblright{}
thicknesses are significantly different depending on the preparation method of the oxide: 5 nm for Hydrogen Silsesquioxane (HSQ) and 30nm for sputtered oxide, for instance.\cite{Anonymous:jlM9QbYX,Stangl:2010bh}
In these works, the existence of a critical oxide thickness on GaAs was explained by the opening of ``craters\textquotedblright{}
in the oxide, either by the reaction of Ga with the substoichiometric oxide and/or due to the desorption of As at GaAs surface temperatures above $500^{\circ}$C. To the best of our knowledge,
there are no reports on the role of oxide on Si substrates in Ga-assisted
growth of GaAs nanowires. One should also note that different types
of oxides have been used for nanowire growth, but no direct comparison and detailed characterization
between different types of oxide has been realized. 
Moreover, it is often observed in the community that the optimized
growth conditions for obtaining GaAs nanowires can strongly fluctuate
by changing wafer batches and providers, despite identical nominal properties.
In this work, we investigate the role of oxide in the Ga-assisted growth of GaAs on Si substrates and provide a method for reproducible nanowire fabrication as a function of the surface and oxide characteristics. The different types of oxides are distinguished as a function of stoichiometry, surface roughness, total thickness and processing parameters.

\section{Experimental Details}

GaAs nanowires have been grown by Ga-assisted self-catalyzed method
on Si(111) 2-inch wafers RCA treated from Siltronix and Virginiatech
($10-20$ $\Omega\cdot cm$). The growth was performed in a Molecular Beam Epitaxy machine (MBE) with solid state sources (DCA P600). Previous to growth and in order to ensure a clean surface, all substrates were annealed at $600{^{\circ}}C$ in a separate UHV chamber; such a process is called ``degassing''. The effect of this step on oxide chemistry, thickness and roughness is presented later on. After this step, samples were moved to the growth chamber by robot arm, always in UHV.

The substrates were prepared with different types of oxides: thermal, native and Hydrogen Silsesquioxane (HSQ). Thermal oxide was produced by means of dry oxidation in a Centrotherm furnace at $950^\circ$C in a cleanroom environment. The native oxide was obtained by natural exposure of the Si wafers to air. HSQ oxide was obtained by spinning a HSQ:MIBK solution (XR-1541-002, Dow Corning) at 6000 rpm and annealing them for 5 minutes at $180^\circ$C for removal of the solvent. Without diluting the solution, the oxide thickness achieved was of $28-30nm$; by diluting it (1:4-1:8) thinner oxides were obtained ($8-4nm$). The films were transformed into Silicon dioxide by annealing them at $475{^{\circ}}C$ in $N_{2}$ atmosphere for 1 hour. The solutions were spun on oxide-free Si wafers to avoid the presence of the native oxide. The oxide thickness was controlled by chemical etching with a $NH_{4}F:HF$ $(500:1)$ solution, calibrating the etching rate for every type of oxide used. The oxide thickness was measured with spectroscopic ellipsometry (Sopra GES 5E) and confirmed by Atomic Force Microscopy (AFM) on etched steps. Attenuated total reflection (ATR) IR spectroscopy (Jasco FT/IR 6300 with Pike MIRacle holder) was realized for the characterization of the oxide stoichiometry, by scanning in
the $650-4000cm^{-1}$ range with 100 accumulation. Although, since
the intensity of the signal-to-noise ratio above $1500$cm$^{-1}$ is
extremely low, only the low range ($650-1500 cm^{-1}$) is considered
and reported. Finally, AFM (Bruker) was also used for
the determination of the surface roughness. In the case of completely etched oxides, the substrates were immersed in an isopropanol bath immediately after etching, and then dried under Nitrogen flow just before loading in the UHV environment. The conditions under which growths were performed were the following: the
substrate temperature ranged between $580{^{\circ}}C$ and $660{^{\circ}}C$;
such values were measured by means of a calibrated pyrometer. The Ga rates used
were between $0.25 A/s$ and $1.25 A/s$; as calibrated on planar growth by means of Reflection High Energy Electron Diffraction (RHEED). The As fluxes
were from $2.5*10^{-6} torr$ to $4.9*10^{-6}torr$; the flux
was calibrated by means of a beam flux monitor gauge. Scanning Electron
Microscopy (SEM) was used for the morphological characterization of the samples.

\section{Experimental Results}

\subsection{Chemical composition of the oxides\label{subsec:ChemCompOxide}}

\begin{table}[t]
\begin{centering}
\caption{Characteristic phononic modes of silicon oxide.\label{tab:Characteristic-phononic-modes} }

\par\end{centering}

\centering{}%
\begin{tabular}{ccc}
\hline 
Phononic Mode  & Position $(cm^{-1})$  & Ref.\tabularnewline
\hline 
\hline 
$Si-O-Si$  & 1107  & [\onlinecite{Boyd:1982kn,Pai:fz,Kim:2003br,Ono:1998jf}] \tabularnewline
\hline 
TO $SiO_{2}$  & 1000-1150  & [\onlinecite{Tian:2010iz,Weldon:1999hf,Queeney:2000km}]\tabularnewline
\hline 
LO $SiO_{2}$  & 1200-1250  &[ \onlinecite{Anonymous:cDm1LXIV,Tian:2010iz,Weldon:1999hf,Queeney:2000km}]\tabularnewline
\hline 
\end{tabular}
\end{table}

We start by listing and comparing the nature of the various oxides used in this work. Thermal oxide is a mostly stoichiometric oxide ($SiO_{2}$),which can be produced by oxidation of Silicon at $800-1200{^{\circ}}C$ under a controlled Oxygen flux; it exhibits low roughness ($\sim>0.6 nm$).
\cite{Ohmi:ihawLB9H,Tian:2010iz,Anonymous:qSDNqIRs,Muller:4ei7Orkm}
Native oxide is a thin layer of oxide formed by the natural exposure of a Si wafer to air; it follows the surface roughness of the Silicon substrate and it grows monolayer by monolayer.\cite{Neuwald:sk2aXoay,Anonymous:BSAUS+Pe,Anonymous:1kXBocAk,Soria:2012gm}
The chemical composition of native oxide depends on its thickness.
For thicknesses of few monolayers, it mainly consists of $Si-O-Si$. The Oxygen content increases for larger thicknesses, though it remains sub-stoichiometric with respect to thermal oxide. HSQ oxide is obtained by annealing a Hydrogen Silsesquioxane resin on a Silicon wafer previously etched with HF. The thickness can be tuned by the dilution of the resin solution and the spinning rate.\cite{Albrecht:1998iu,Yang:91OsoeUQ} Annealing the HSQ resin at $450{^{\circ}}C$ transforms the cage structure of HSQ monomer into a network, whose chemical composition is
$SiO_{x}$ with $1<x<2$, depending on the annealing temperature.\cite{Albrecht:1998iu,Yang:91OsoeUQ}

\begin{figure}[t]
\includegraphics[width=8.6cm]{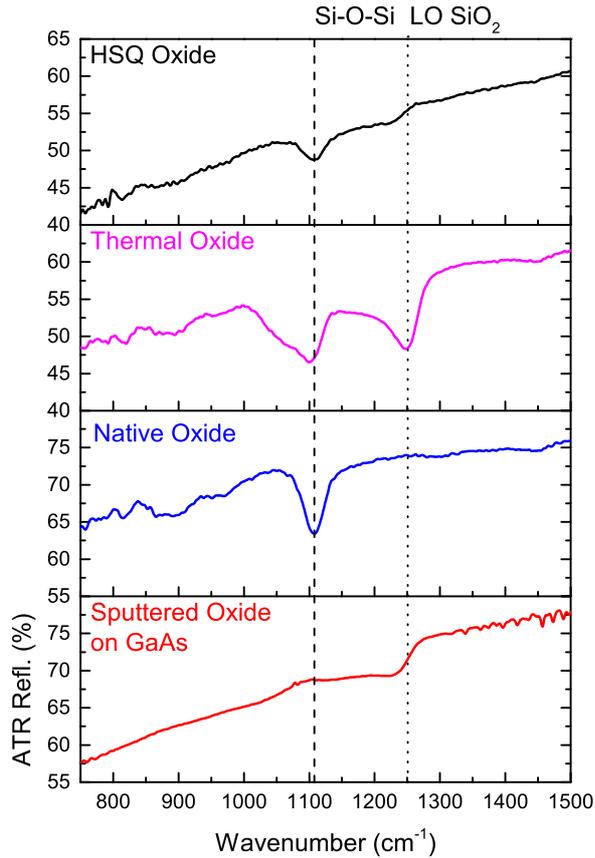}

\caption{\label{fig:IR_DifferentOxides} (Color Online) ATR FTIR spectroscopy of different oxides.\label{fig:ATR-FTIR-spectroscopy}
The peak at $1250 cm^{-1}$corresponds to the LO mode of $SiO_{2}$,
whereas the peak at $1107 cm^{-1}$ is related to the presence of interstitial
oxide $Si-O-Si$. The TO mode of $SiO_{2}$ is located around $1050 cm^{-1}$.
Thermal oxide is the only oxide that shows TO and LO modes of $SiO_{2}$.
HSQ present a downshifted broad peak around $1200cm^{-1}$correspondent
to non-stoichiometric oxide $SiO_{x}$ with $x<2$. All the oxides
show the interstitial oxide peak $(1107 cm^{-1})$ but not the sputtered
oxide on GaAs; this proof that $Si-O-Si$ is peculiar of Silicon-Silicon
oxide interfaces. Similar result was obtained with HSQ oxide on GaAs substrate (see Supplementary Information). }
\end{figure}

The stoichiometry of Silicon oxides are often characterized with Fourier Transform Infrared Spectroscopy (FTIR). The main absorption bands characteristic of Silicon oxides are the interstitial oxygen band $(Si-O-Si)$, centered at $1107cm^{-1}$, \cite {Boyd:1982kn,Pai:fz,Kim:2003br,Ono:1998jf} the transverse optical phonon (TO) of $SiO_{2}$ around $1000cm^{-1}$,\cite{Tian:2010iz,Weldon:1999hf,Queeney:2000km} as well as the longitudinal optical phonon (LO) around $1250cm^{-1}$, \cite{Anonymous:cDm1LXIV,Tian:2010iz,Weldon:1999hf,Queeney:2000km} as reported in table \ref{tab:Characteristic-phononic-modes}. 
Suboxides of the form $SiO_{x}$ with $1\leq x<2$ are characterized by an absorption band downshifted and broadened with respect to the TO $SiO_{2}$. The shift can be related to the stoichiometry in an approximate manner $x\rightarrow1$.\cite{Queeney:2000km} Examples of ATR-FTIR spectra obtained from the different oxides are shown in Fig. \ref{fig:IR_DifferentOxides}.

The $Si-O-Si$ absorption band is observed for all the oxides (thermal, native and HSQ), albeit with different intensities. On the other hand the presence of  additional absorption band depends on the oxide type: 

\begin{itemize}
	\item Thermal oxide shows a clear $LO-SiO_{2}$ positioned at $1250cm^{-1}$, indicating it is stoichiometric oxide; 
	\item HSQ oxide show a downshifted additional absorption band ($~1226cm^{-1}$), indicating it is substoichiometric oxide $SiO_{x}$with$(1\leq x<2)$;\cite{Queeney:2000km} 
	\item Native oxide does not show any additional absorption band.
\end{itemize}

Finally, in order to demonstrate that the interstitial oxide is characteristic only of the Silicon - Silicon oxide interface, the spectra of sputtered
oxide on GaAs is shown (see Fig. \ref{fig:IR_DifferentOxides}); the same result has been achieved with HSQ on GaAs (see Supplementary Information). No absorption band is observed at $1107cm^{-1}$, showing that no interstitial oxide is present if the interface is GaAs - Silicon oxide. On the other hand a broad absorption is verified
around $1226cm^{-1}$, characteristic of sub-stoichiometric oxide. In the following, FTIR spectra of substrates prior to growth are going to be used to understand the role of the surface chemistry on the growth mechanisms of Ga-assisted growth of GaAs nanowires by MBE.

\subsection{Thermal Oxide\label{sec:Thermal-Oxide}}

\begin{figure}[t]
\includegraphics[width=8.6cm]{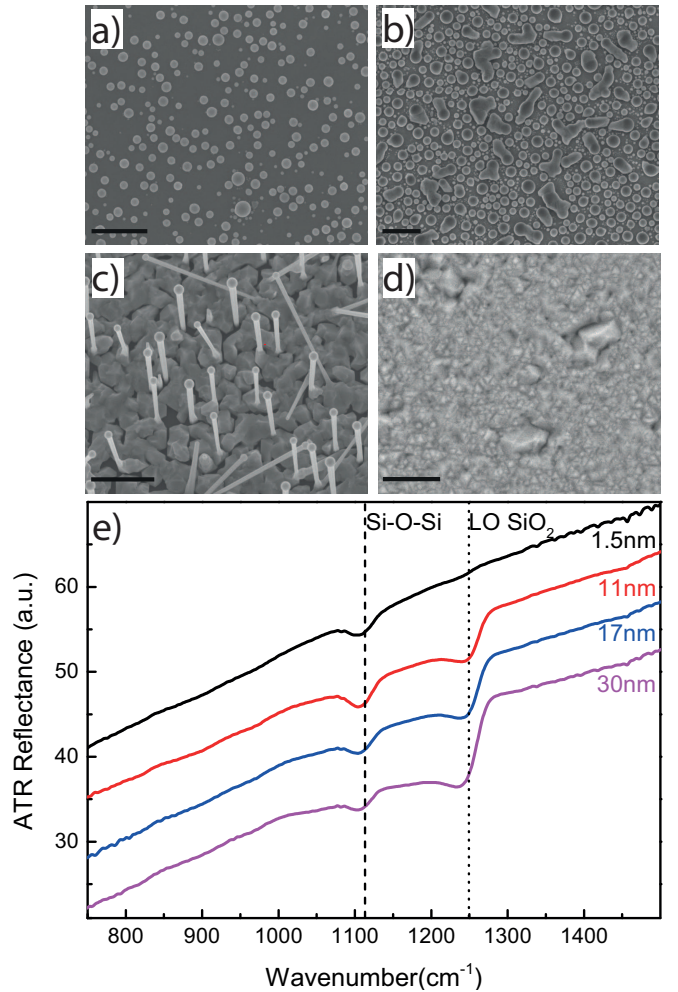}

\caption{\label{fig:ThermalOxideSEM_IR1} (Color Online) Growth of GaAs nanowires on Si $(111)$
substrates covered by thermal oxide. The same substrate has been etched
with $NH_{4}F:HF$ (500:1) SEM micrographs of GaAs nanowires growth
attempt on (a) $24$ nm, (b) $10$nm and (c) $1.5$nm thick thermal
oxide. In (d) growth without oxide is reported. Growth has been performed
simultaneously for all the different oxide thicknesses. In (e) ATR-FTIR
spectra of thermal oxide with different thicknesses are reported.
By decreasing the oxide thickness the $LO$ $SiO_{2}$ is decreasing
in intensity, whereas the $Si-O-Si$ remain unchanged. Even at low
thicknesses ($\sim1-2nm)$ the $LO$ $SiO_{2}$ could be observed,
showing that in thermal oxide the compositional gradient from $Si-O-Si$
to $SiO_{2}$ is sharp. The scale-bar corresponds to $1\mu m$.}
\end{figure}

We start by the study of Ga-assisted growth of GaAs nanowires on thermally oxidized Silicon substrates. We first show the effect of the oxide thickness. For this, we prepared substrates with four different oxide thicknesses ($30$ down to $0$ nm) by a combination of optical lithography and etching (see Supplementary Information). This allowed us to investigate simultaneously several oxide thicknesses under identical experimental conditions.\\
 Fig. \ref{fig:ThermalOxideSEM_IR1} shows SEM micrographs of substrates with varying oxide thicknesses after performing the same growth process. The growth conditions were substrate temperature $T=600^{o}C$, Gallium rate $Ga=1.25A/s$ and Arsenic beam flux pressure $As=2.5*10^{-6}$torr). Under these conditions nanowire growth was observed only for an oxide thickness between 1 and 2 nm ( Fig.\ref{fig:ThermalOxideSEM_IR1}(c)). Similar thickness selectivity results were obtained under other conditions leading to nanowire growth.
In the case of thicker oxides, Ga droplets were observed on the surface (see Fig. \ref{fig:ThermalOxideSEM_IR1}(a)-(b)). For oxide-free Silicon surfaces, textured two-dimensional growth was found (see Fig.\ref{fig:ThermalOxideSEM_IR1}(d)).
The question here is what makes $1-2nm$ thermal oxide so prone for Ga-assisted nanowire growth. In order to shed some light to this question the chemical composition and surface roughness of the thermal oxide with different thicknesses were investigated. In Fig.\ref{fig:ThermalOxideSEM_IR1} (e) the ATR-FTIR spectra of the thermal oxide for different thicknesses are shown. It is interesting to note that the intensity of the LO $SiO_{2}$ is proportional to the
thickness, while the interstitial $Si-O-Si$ mode exhibits the identical amplitude for
all thicknesses. This corroborate the interfacial nature of the interstitial oxide $Si-O-Si$, as mentioned in subsection \ref{subsec:ChemCompOxide}.  Additionally, for oxide thicknesses around $1-2nm$,
the ATR-FTIR spectra is composed only by the interstitial band. As reported
by Muller et al, \cite{Muller:4ei7Orkm} at the interface with
Silicon and for about 1 to 2 nm thermal oxide is composed mainly by $Si-O-Si$. In accordance to our observation stoichiometric Silicon dioxide appears only at larger
thicknesses. Thus it seems that interstitial oxide must be more prone to nucleation of nanowires by the Ga-assisted method. The role of the chemical nature of the surface for successful nanowire growth will be further elaborated by examining growth on oxides with different stoichiometry here below.


\begin{table*}[ht]
\begin{centering}
\caption{Thickness and roughness of different types of oxides before and after
the pre-growth annealing of 2 hours. The measurements have been performed first by ellipsometry, and then confirmed by cross-section TEM.\label{tab:Thickness-and-Roughness}}

\par\end{centering}

\centering{}%
\begin{tabular}{lccccc}
\hline 
\multirow{2}{*}{Oxide Type } & \multirow{2}{*}{Annealing $(^{o}C)$ } & \multicolumn{2}{c}{Thickness (nm) } & \multicolumn{2}{c}{Roughness RMS (nm) }\tabularnewline
\cline{3-6} 
 &  & Before  & After  & Before  & After \tabularnewline
\hline 
\hline 
Native Oxide (Virginia)  & 600  & 0.9$\pm$0.6  & 0.8$\pm$0.6   & 0.3$\pm$0.5  & 0.5$\pm$0.5 \tabularnewline
\hline 
Native Oxide (Siltronix)  & 600 & 2.3$\pm$0.6  & 2.1$\pm$0.6  & 0.8$\pm$0.5  & 5.3$\pm$0.5 \tabularnewline
\hline 
Thermal Oxide  & 600 & 1.4$\pm$0.6  & 1.2$\pm$0.6   & 3.4$\pm$0.5  & 1.3$\pm$0.5 \tabularnewline
\hline 
HSQ Etched  & 400 & 9.4$\pm$0.9  & 8.8$\pm$0.9   & 1.1$\pm$0.5  & 3.1$\pm$0.5 \tabularnewline
\hline 
HSQ As Spun  & 400  & 8.1$\pm$0.6  & 8.1$\pm$0.6  & 3.6$\pm$0.5  & 1.2$\pm$0.5 \tabularnewline
\hline 
\end{tabular}
\end{table*}

\begin{figure}[t]
\includegraphics[width=8.6cm]{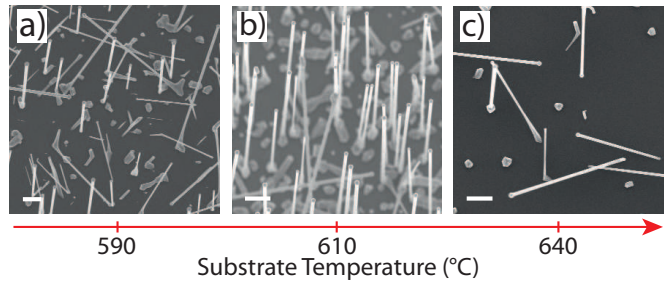}

\caption{\label{fig:ThOxTempSeries} (Color Online) SEM micrographs of GaAs growth on Si(111)
substrate covered with $1-2nm$ thick thermal oxide. The growth conditions were with
identical Ga rate and As flux ( $Ga=0.75A/s$, $As=3.9*10^{-6}torr$
), but different substrate temperature (increasing from left to right).
At $590^\circ$C nanowires growth is achieved, but with
low yield and parasitic growth. By increasing temperature to $610{^\circ}C$
the yield is increased; the same trend is observed up to $625^\circ$C
(figure not reported). A further increase of temperature up to $640^\circ$C
show a complete loss of nanowire orientation. The scale-bar corresponds to $1\mu m$.}
\end{figure}

It must be borne in mind that prior to growth degassing is performed; such a step might affect the surface morphology and composition. Therefore oxide thickness and surface roughness before and after degassing were investigated. The values are reported in Tab. \ref{tab:Thickness-and-Roughness}.
While the oxide thickness measured by spectroscopic ellipsometry remains constant before and after annealing, the surface roughness is reduced (from $3.4$ to $1.3nm$). The reduction of surface roughness might come from minimization of surface energy, and ``glass'' flow during degassing. After having found the ideal oxide thickness for growth, we explore the parameter space for nanowire growth in terms of temperature and fluxes. Fig. \ref{fig:ThOxTempSeries} shows the effect of the substrate temperature on nanowire growth: by increasing temperature from $590^\circ$C to $610^\circ$C (Fig. \ref{fig:ThOxTempSeries}(a)-(b)) an increase in nanowire density and yield of growth perpendicular to the substrate is observed, coherently with what reported by Krogstrup
et al and Giang et al. \cite{Giang:2012tx,Krogstrup:2010ux,RussoAverchi:2012rZ}
By further increasing the growth temperature to $625^\circ$C
the nanowire density decreases (figure not reported). Surprisingly, a further increase in the substrate temperature results in a significant ratio of nanowires misoriented with respect to the substrate normal (see
Fig. \ref{fig:ThOxTempSeries} (c)).

\begin{figure}[t]
\includegraphics[width=8.6cm]{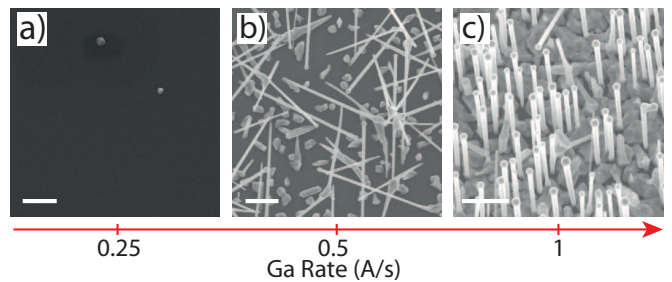}

\caption{\label{fig:ThOxGaSeries} (Color Online) SEM micrographs of GaAs growth on Si(111)
substrate covered with thermal oxide. The growth conditions were with identical substrate temperature and As flux ($T=630{^\circ}C$, $As=2.5*10^{-6}torr$), but different Ga rate (increasing from left to right). At low Ga rate ($0.25A/s$) no nanowire growth is achieved (a). By increasing
the rate to $0.5A/s$ nanowires are obtained, together with
parasitic growth (b). Further increase in Ga rate led to an increase of the vertical yield, together with a further increase of parasitic growth. The scale-bar correspond to $1\mu m$.}
\end{figure}

The effect of Ga rate on the morphology of nanowires is shown in Fig. \ref{fig:ThOxGaSeries}. At
low Ga rate ($0.25A/s$) nanowires do not grow (Fig. \ref{fig:ThOxGaSeries}(a)).
When the rate is increased up to $0.5-0.7A/s$, growth is achieved
(see Fig. \ref{fig:ThOxGaSeries}(b)). A further increase of Ga rate
results in an increase of nanowire density together with the occurrence
of parasitic growth (Fig. \ref{fig:ThOxGaSeries}(c)).


We would like to add a comment on the etching process of oxidized
silicon wafers for obtaining a thin thermal oxide. Even though the wet etching process was realized in a controlled manner, it happens to be inhomogeneous across the wafer. In order to obtain a better homogeneity across the sample, we started with wafers with a thinner thermal oxide (e.g. $5 nm$ instead of $30 nm$). This results in a much better homogeneity across the sample.

\subsection{Native Oxide\label{sec:Native-Oxide}}

\begin{figure}[b]
\includegraphics[width=8.6cm]{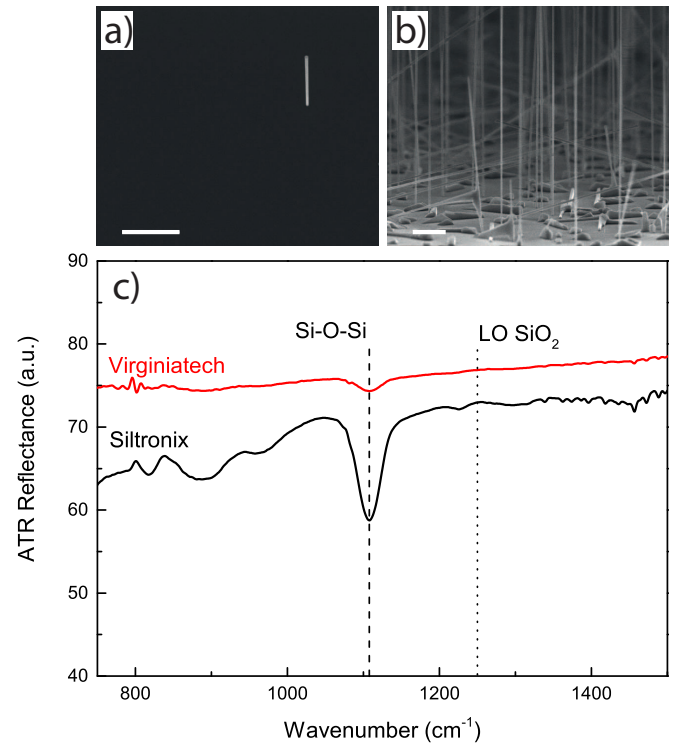}

\caption{\label{fig:NativeOxideSEM_IR} (Color Online) Native oxide grown on wafers of different
providers (Siltronix, Virginiatech) with same doping concentration
and same surface treatment (RCA). Growth has been performed with same
process parameters ($T=610^{o}C$, $Ga=0.5A/s$, $As=2.5*10^{-6}torr$),
and in the case of Virginiatech no growth was achieved (a), whereas
in the case of Siltronix growth was achieved (b). (c) ATR-FTIR spectra
of Siltronix and Virginiatech native oxides. Siltronix shows stronger
absorption band of $Si-O-Si$ compared to Virginiatech at comparable
thickness and roughness (see Tab. \ref{tab:Thickness-and-Roughness}).The
scale-bar corresponds to $2\mu m$.}
\end{figure}

We turn now our attention towards the growth on Silicon substrates presenting only a native oxide. As the native oxide thickness is in the order of few nanometer, \cite{Soria:2012gm,Anonymous:BSAUS+Pe}
we did not perform any study on the ideal native oxide thickness.
We just kept the factors affecting the thickness constant: doping concentration, surface cleaning process and surface orientation.\cite{Anonymous:BSAUS+Pe}
For this reason we used and compared (111) Si substrates with the same nominal resistivity, but delivered from two different providers.
Nanowire growth was obtained for both types of wafers but in very different growth conditions. As an example, in Fig. \ref{fig:NativeOxideSEM_IR}
we show SEM images from two growths performed under identical conditions
($T=610^{o}C$, $Ga=0.5A/s$, $As=2.5*10^{-6}torr$), but on Si wafers from two different providers ((a) Virginiatech and (b) Siltronix). A dense array of nanowires is obtained on Siltronix wafers (see Fig. \ref{fig:NativeOxideSEM_IR}(b)),
while an extremely low density of nanowires is observed on the Virginiatech (see Fig. \ref{fig:NativeOxideSEM_IR}(a)). By comparing the ATR-FTIR spectra of both wafers (\ref{fig:NativeOxideSEM_IR}
(c)) we observe a stronger presence of interstitial $Si-O-Si$ oxide in the case of Siltronix (see Fig. \ref{fig:NativeOxideSEM_IR}).
We also find that this native oxide is slightly thicker and rougher 
(see Tab. \ref{tab:Thickness-and-Roughness}): $2.3$ versus $0.9nm$
for the thickness and $0.9$ versus $0.3nm$ for the roughness.


\begin{figure}[t]
\begin{centering}
\includegraphics[width=8.6cm]{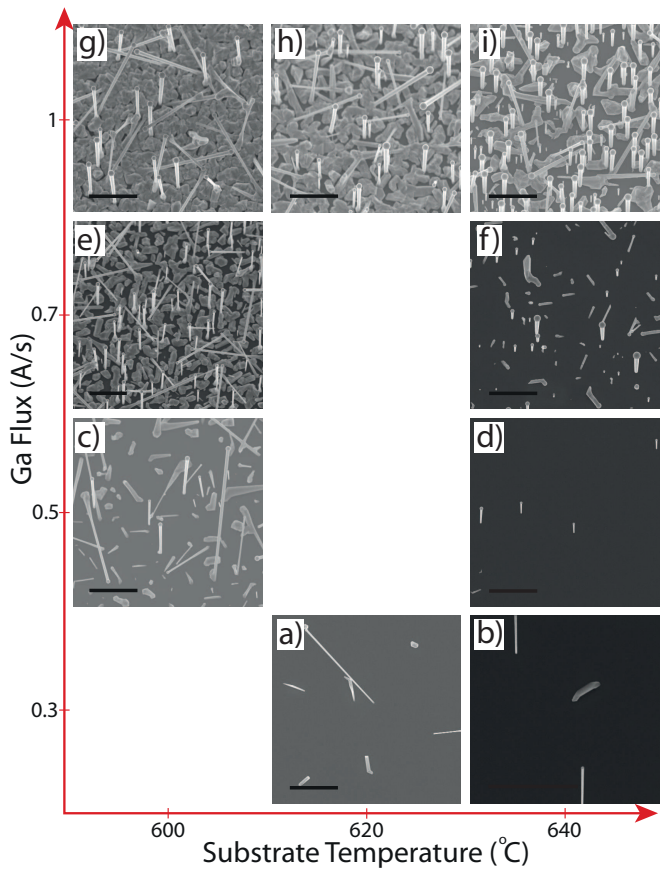} 
\par\end{centering}

\caption{\label{SEM_NatOxVirg_Ga_VS_T} (Color Online) SEM micrographs of GaAs nanowires grown on
Virginiatech wafers covered with native oxide. The As pressure used in all the growths shown is constant at $2.5*10^{-6}torr$. By moving
from bottom to top Ga flux increases, whereas from left to right the
substrate temperature increases. At low Ga flux and substrate temperatures
(a)-(b) low material deposition is observed. On the other hand
the higher the Ga flux (c)-(e)-(g), the more material is deposited
on the surface, producing nanowires of various orientations and polycrystalline
parasitic growth. If also the temperature is increased the density
of vertical NWs strongly increases (g)-(h)-(i), regardless the low
V/III ratio. Growth at temperature above $640^\circ$C
was also attempted, but no growth was observed anymore. The scale-bar
corresponds to $2\mu m$.}
\end{figure}

In order to understand whether nanowire growth was possible for wafers
exhibiting a smaller amount of interstitial oxide (Virginiatech),
we varied extensively the growth parameters. We varied both Ga rate
and substrate temperature respectively from $0.3A/s$ to $1A/s$,
and from $600^\circ$C to $660^\circ$C.
The results obtained are shown in Figure \ref{SEM_NatOxVirg_Ga_VS_T}: at low Ga flux and substrate temperatures ((a)-(b)) no growth of
vertical wires is observed. On the other hand the higher the Ga flux
((c)-(e)-(g)), the more material is deposited on the surface, resulting
in growth of nanowires of various orientations and polycrystalline
parasitic layer. By increasing the temperature, the density of nanowires
growing perpendicularly to the surface increases significantly ((g)-(h)-(i)),
regardless the low V/III ratio. Growths at temperature above $640^\circ$C
were also attempted, but no nanowires were observed. Increasing substrate
temperature the occurrence of parasitic growth decreases (see Fig.
\ref{SEM_NatOxVirg_Ga_VS_T} (g)-(h)-(i), or (e)-(f)). This effect,
coupled with an increase of Ga rate led to successful nanowire growth. \\


 It must be said that for native oxide we did not observe any variation in the oxide thickness nor chemical composition  by degassing (see Tab. \ref{tab:Thickness-and-Roughness}).
However, in the case of Siltronix Si wafers the surface roughness
increased from $0.8$ to $5.3nm$.

\subsection{HSQ Oxide\label{sec:HSQ-Oxide}}

Finally, we looked at the nanowire growth on Silicon substrates covered with HSQ. We started our study by optimizing the oxide thickness.
Substrates with four different oxide' thicknesses on the same wafer were prepared (details on the method are provided in Supplementary Information). Several growths were performed at the same time on oxide thicknesses ranging from $2nm$ up to $24nm$. A degassing temperature of $400^\circ$C was used in the case of HSQ oxides, since no growth was achieved with $600^\circ$C. We suspect that degassing at higher temperature lead to the formation of a compact oxide that does not allow nucleation.\\

\begin{figure}[t]
\includegraphics[width=8.6cm]{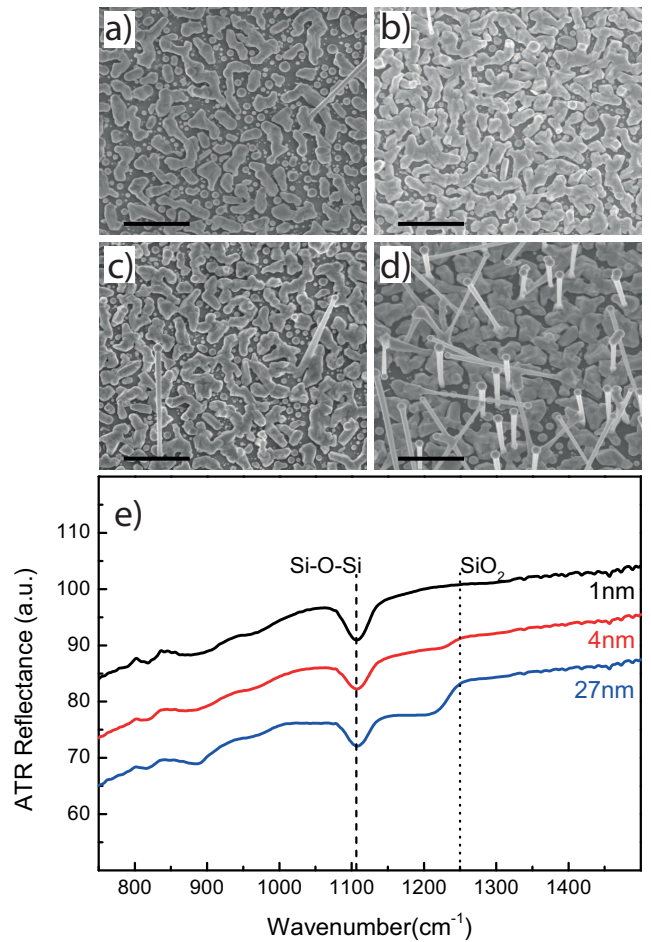}

\caption{\label{fig:HSQOxideSEM_IR} (Color Online) Growth of GaAs nanowires on Si (111) substrates
covered with HSQ oxide. Thickness was controlled by etching down the
oxide with $NH_{4}F:HF$ diluted $500:1$. In (a)-(b)-(c) and (d)
the SEM micrographs show the growth attempts on respectively $19-15-8$
and $5nm$ oxide thicknesses. Only in the latter case growth was performed
successfully; the growth conditions were substrate temperature $595^\circ$C,
Ga rate of $1A/s$ and As flux of $2.5*10^{-6}torr$. (e) ATR-FTIR
spectra of HSQ oxide with different thicknesses: the absorption band
of $SiO_{x}$ decreases in intensity by decreasing oxide thickness.
On the other hand the absorption band of $Si-O-Si$ does not decrease
in intensity, showing that it is related to the Silicon-Silicon oxide interface.The
scale-bar corresponds to $2\mu m$.}
\end{figure}

\begin{figure}[tr]
\includegraphics[width=8.6cm]{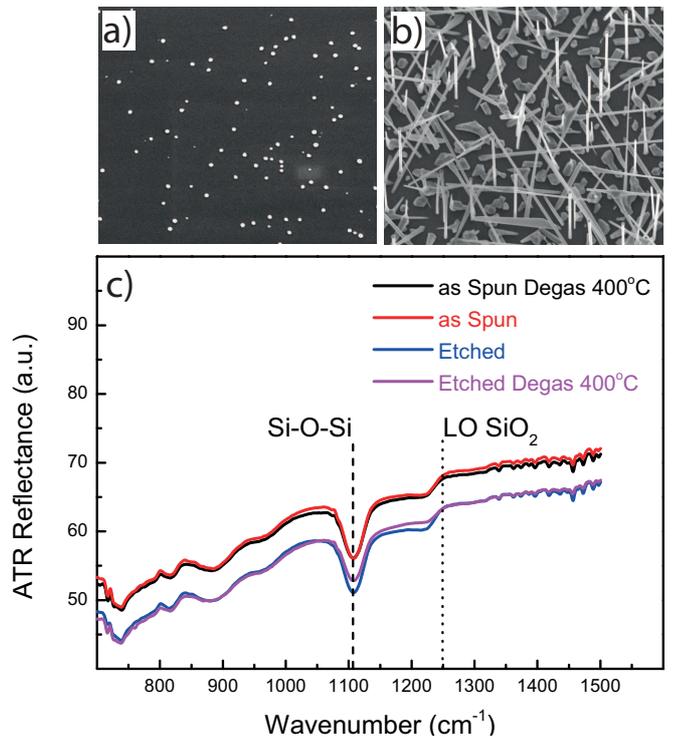}

\caption{\label{fig:HSQEtchedVSasSpun} (Color Online) HSQ oxide SEM micrographs of the growth
attempts on oxide prepared by (a) direct dilution, called ``as spun,''
or (b) etch from a more concentrated solution, called ``etched''.
Nanowires growth was successful only in the case of the etched HSQ.
In (c) is shown the evolution of the ATR-FTIR spectra before and after
the annealing step previous to growth: in the case of the \emph{as
spun} no variation is observed, whereas in the case of the \emph{etched}
a decrease of intensity of the absorption bands of $LO$ $SiO_{2}$
and of $Si-O-Si$ was observed. Such a decrease observed in the IR
spectra is supported by an observed decrease in oxide thickness (see
Tab. \ref{tab:Thickness-and-Roughness}).}
\end{figure}

 The result of this investigation is reported in Fig.\ref{fig:HSQOxideSEM_IR}:
(a)-(b)-(c) and (d) correspond respectively to oxide thicknesses of
$19-15-8$ and $5nm$. Growth was performed at a substrate temperature
of $595^\circ$C, Ga rate of $1A/s$ and As flux of $2.5*10^{-6}torr$.
Here, the critical thickness for nanowire growth is higher than in
the case of thermal oxide, and comparable to what has been reported
for HSQ on GaAs.\cite{Rieger:UY13hl5a} Growth of nanowires was observed only when
the thickness of the oxide was below $5-6nm$, as shown in \ref{fig:HSQOxideSEM_IR}
(d). In Fig. \ref{fig:HSQOxideSEM_IR} (e) the ATR-FTIR spectra of the
HSQ oxide for different thicknesses are shown. The intensity of the
LO $SiO_{x}$ decreases with the thickness, while the spectral position
is maintained. As in the case of thermal oxide, we observe that the interstitial $Si-O-Si$
mode is maintained in intensity, corroborating the interfacial nature of interstitial oxide. Additionally, for oxide thicknesses of about $4nm$, the ATR-FTIR spectra exhibit mainly the interstitial
band, and the LO $SiO_{x}$ is shifted. This means that the composition of
HSQ is not homogeneous across the thickness.\cite{Albrecht:1998iu,Yang:91OsoeUQ,Queeney:2000km} The oxide reduced to
a thickness of about $5-6nm$ is formed mainly by interstitial oxide
$Si-O-Si$ with a scarce proportion of $SiO_{x}$, since most of the
$SiO_{x}$ has been etched away. In order to understand if the thickness
is the only determining parameter for growth on HSQ oxide, we prepared different
HSQ layers with the same thickness but by different processing: 
\begin{itemize}
\item We spun HSQ from a  MIBK diluted solution (1:8) to form
directly the desired thickness and avoid etching. This type of sample
will be called ``as spun''(see Fig. \ref{fig:HSQEtchedVSasSpun} (a)). 
\item We spun HSQ from non-diluted solution, and etched it down to the
desired thickness by means of wet etch with $NH_{4}F:HF$ $500:1$.
In the following, this type of sample will be called ``etched down''(see Fig. \ref{fig:HSQEtchedVSasSpun}.
(b)). 

\end{itemize}


In order to understand the different behavior of the HSQ, oxide thickness,
roughness and chemical composition were measured before and after the
pre-growth annealing(see Tab. \ref{tab:Thickness-and-Roughness}).
In fact the substrates prepared in two different manner evolve in an opposite way. While the  HSQ etched down shows a reduced thickness and increased roughness upon annealing,
the other oxide keep constant the thickness and become smoother.
The chemical composition is observed to be identical for both the
etched and as spun oxides, being unchanged before and after the pre-growth
annealing (see Fig. \ref{fig:HSQEtchedVSasSpun} (c)).

\section{Discussion}

\begin{figure}[t]
\includegraphics[width=8.6cm]{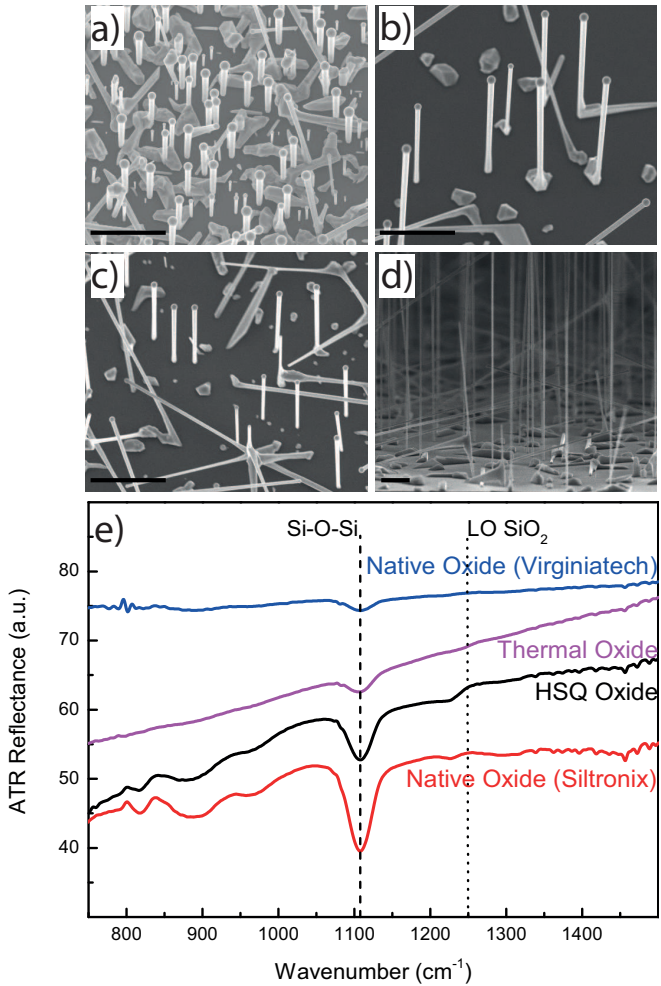}

\caption{\label{fig:DifferentOxideConclusions} (Color Online) GaAs nanowires growth on Si
(111) with different Silicon oxides. SEM micrographs of (a) native
oxide (Virginiatech), (b) thermal oxide, (c) HSQ oxide and (d) native
oxide (Siltronix) are reported. In (e) the ATR-FTIR spectra of the
different oxides are shown: native oxide (Virginiatech) is the oxide
that shows lower interstitial oxide absorption band, then thermal,
HSQ and native oxide (Siltronix) show progressively higher content.The
conditions to achieve growth on the different oxides were observed
to be different: in the case of native oxide (Virginiatech) the conditions
were $Ga=1A/s$ and $T_{sub}=640^\circ$C, for thermal
oxide $Ga=0.7A/s$ and $T_{sub}=630^\circ$C, for HSQ
were $Ga=0.5A/s$ and $T_{sub}=610^\circ$C and native
oxide (Siltronix) were $Ga=0.2A/s$ and $T_{sub}=610^\circ$C.
A correlation between interstitial oxide presence and growth conditions
was observed: the lower the content of interstitial oxide, the higher
the Ga rate and substrate temperature. The scale-bars correspond to
$2\mu m$.}
\end{figure}

\begin{figure}[t]
\includegraphics[width=8.6cm]{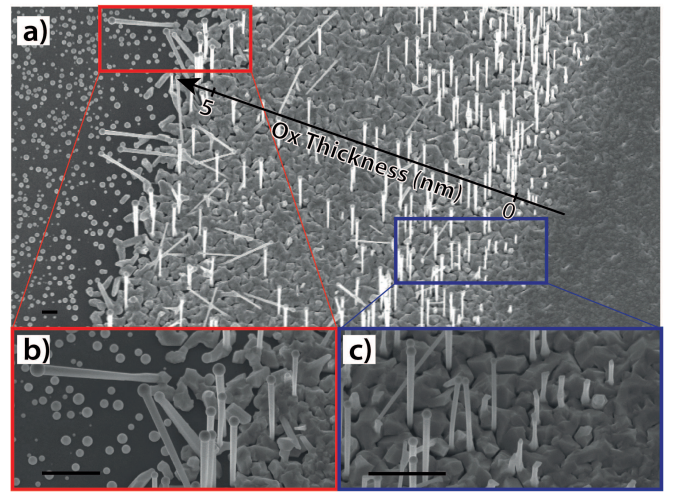}

\caption{\label{fig:thermalOxideThicknessgradientSEM} (Color Online) SEM micrographs of GaAs
growth on Si(111) substrate covered with thermal oxide thickness range:
in (a) from right to left oxide thickness goes from $0nm$ to $5nm$.
If thickness is homogeneous around $5nm$ (b) no growth could be achieved.
On the other hand by decreasing oxide thickness vertical and tilted
nanowires start to appear. (c) When thickness is lowered even more,
only vertical nanowires appear, until only two dimensional growth
is observed. }
\end{figure}

In this section we discuss the results concerning the success and/or optimal nanowire growth conditions as a function of the oxide chemistry (stoichiometry) and surface roughness. Oxide chemistry range from a composition of $SiO_2$ in the case of thermal oxide, $SiO_x$ $(1<x<2)$ for HSQ, down to $SiO_{0.5}$ in the case of interstitial oxide. Therefore the overall range of composition can be described by $SiO_x$ with $x$ ranging from 0 to 2. For simplicity sake, we start considering the two extreme cases: growth on stoichiometric $SiO_{2}$ ($x=2$) and growth on an oxide-free silicon ($x=0$). In fact, in both cases no successful GaAs nanowire growth was obtained in any of the conditions used. Nanowire growth could only be achieved when the thermal oxide was around $2nm$ thick. As shown by the FTIR study, at this thickness the oxide is not stoichiometric $SiO_2$. This sets some kind of ``boundary conditions'' for the growth of self-catalyzed GaAs nanowires on Si substrates (see Fig. \ref{fig:thermalOxideThicknessgradientSEM}). In order to achieve growth, a sub-stoichiometric silicon oxide is needed ($SiO_{x}$ with $0<x<2$). The exact nature of this sub-stoichiometric oxide has also a direct effect on the growth conditions needed for growth: for example, in the case of interstitial oxide $Si-O-Si$ ($x=0.5$) growth of GaAs nanowires was always observed. A correlation between the interstitial oxide content regardless of the oxide and the conditions to achieve growth was found. The lower the $Si-O-Si$ content, the higher the substrate temperature and
Ga rate were needed to achieve growth (see Fig. \ref{fig:DifferentOxideConclusions}).
As an example, in the case of Virginiatech Si wafers, (see Fig. \ref{fig:DifferentOxideConclusions}(a)) the presence of interstitial oxide was weakest. The conditions to achieve growth were of $Ga=1A/s$ and $T_{sub}=640^\circ$C. Thermal oxide presented the second lowest $Si-O-Si$ content (see Fig. \ref{fig:DifferentOxideConclusions}(b)).
Growth was achieved with $Ga=0.7A/s$ and $T_{sub}=630^\circ$C.
HSQ has a stronger $Si-O-Si$ absorption band compared to thermal oxide, and the conditions for successful growth were $Ga=0.5A/s$ and $T_{sub}=610^\circ$C (see Fig. \ref{fig:DifferentOxideConclusions}(c)).
The oxide that showed stronger interstitial oxide absorption band
was native oxide was the Siltronix Si wafer. In this case, growth was achieved with even lower Ga rate and substrate temperature ($Ga=0.27A/s$ and $T_{sub}=610^\circ$C), as shown in Fig. \ref{fig:DifferentOxideConclusions}(d).
On the other hand in the case of sub-oxides $SiO_{x}$ ($1<x<2$)  nanowire growth was possible, but strongly dependent on surface roughness:
only higher surface roughness ($>3nm$) lead to nanowire growth(see Fig.\ref{fig:HSQEtchedVSasSpun}). 

Finally, our results raise new questions which should be addressed for a better understanding of the Ga-assisted growth of GaAs nanowires on silicon:
a) what is the role of $Si-O-Si$ in the nucleation of nanowires?
Why does $SiO_{2}$ not work in the same manner? b) what is the role
of surface roughness in the nucleation of nanowires? c) why do growth
conditions must be tuned to high Ga rates when the surface roughness
is low? We believe several elements should be considered: interstitial
oxide might just be more likely to react with Ga impinging on the
surface, reaction points being pinning sites for the nucleation of a Ga droplet leading to growth. Surface roughness might play also a role in the
pinning of Ga droplets and precipitation of GaAs from the Ga droplet.
Still, more studies need to be performed to confirm this and provide
further understanding on the microscopic model of nanowire growth.

\section{Conclusions}

In conclusion, we have shown that the nature of oxide is a key parameter for obtaining successful GaAs nanowire growth on Si substrates by the Gallium-assisted method. We show that interstitial oxide $Si-O-Si$ is very favorable to nanowire growth. The process window is related to the amount of oxide, the amount of $Si-O-Si$ content being directly linked to the Ga rate and substrate temperature needed. By increasing the Oxygen content, the process window is decreased and depends strongly on the surface roughness. In agreement with all this, we found different critical thicknesses of the oxide for successful nanowire growth: $1-2nm$ for thermal and native oxide, $5-6nm$ for HSQ. Finally, we could not obtain growth on stoichiometric $SiO_{2}$, or in the complete absence of oxide (see Fig. \ref{fig:thermalOxideThicknessgradientSEM}). This work opens new perspectives for the reproducible integration of GaAs nanowires on silicon.

\section*{ACKNOWLEDGEMENTS}

This work has been funded by ERC Starting Grant 'UpCon', ERANET-RUS InCoSiN,
FP7 NanoEmbrace, SNF GRANT NR. The authors thank Martin Heiss for helpful discussions, M. de la Mata and J. Arbiol for cross-section TEM measurements, Holger Frauenrath and Francesco Stellacci for the availability of
the IR-ATR. 

\bibliographystyle{apsrev}
\bibliography{ref_21_d_1}

\begin{thebibliography}{54}
\expandafter\ifx\csname natexlab\endcsname\relax\def\natexlab#1{#1}\fi
\expandafter\ifx\csname bibnamefont\endcsname\relax
  \def\bibnamefont#1{#1}\fi
\expandafter\ifx\csname bibfnamefont\endcsname\relax
  \def\bibfnamefont#1{#1}\fi
\expandafter\ifx\csname citenamefont\endcsname\relax
  \def\citenamefont#1{#1}\fi
\expandafter\ifx\csname url\endcsname\relax
  \def\url#1{\texttt{#1}}\fi
\expandafter\ifx\csname urlprefix\endcsname\relax\def\urlprefix{URL }\fi
\providecommand{\bibinfo}[2]{#2}
\providecommand{\eprint}[2][]{\url{#2}}

\bibitem[{\citenamefont{Czaban et~al.}(2009)\citenamefont{Czaban, Thompson, and
  LaPierre}}]{Czaban:2009qw}
\bibinfo{author}{\bibfnamefont{J.}~\bibnamefont{Czaban}},
  \bibinfo{author}{\bibfnamefont{D.}~\bibnamefont{Thompson}}, \bibnamefont{and}
  \bibinfo{author}{\bibfnamefont{R.}~\bibnamefont{LaPierre}},
  \bibinfo{journal}{Nano Letters} \textbf{\bibinfo{volume}{9}},
  \bibinfo{pages}{148} (\bibinfo{year}{2009}).

\bibitem[{\citenamefont{Colombo et~al.}(2009)\citenamefont{Colombo, Heiss,
  Gratzel, and i~Morral}}]{Anonymous:DMWvgapA}
\bibinfo{author}{\bibfnamefont{C.}~\bibnamefont{Colombo}},
  \bibinfo{author}{\bibfnamefont{M.}~\bibnamefont{Heiss}},
  \bibinfo{author}{\bibfnamefont{M.}~\bibnamefont{Gratzel}}, \bibnamefont{and}
  \bibinfo{author}{\bibfnamefont{A.~F.} \bibnamefont{i~Morral}},
  \bibinfo{journal}{Applied Physics Letters} \textbf{\bibinfo{volume}{94}},
  \bibinfo{pages}{173108} (\bibinfo{year}{2009}).

\bibitem[{\citenamefont{Kelzenberg et~al.}(2010)\citenamefont{Kelzenberg,
  Boettcher, and Atwater}}]{Kelzenberg:2010fa}
\bibinfo{author}{\bibfnamefont{M.}~\bibnamefont{Kelzenberg}},
  \bibinfo{author}{\bibfnamefont{S.}~\bibnamefont{Boettcher}},
  \bibnamefont{and} \bibinfo{author}{\bibfnamefont{H.}~\bibnamefont{Atwater}},
  \bibinfo{journal}{Nature Materials} \textbf{\bibinfo{volume}{9}},
  \bibinfo{pages}{239} (\bibinfo{year}{2010}).

\bibitem[{\citenamefont{Tian et~al.}(2007)\citenamefont{Tian, Zheng, Kempa, and
  Lieber}}]{Tian:2007kl}
\bibinfo{author}{\bibfnamefont{B.}~\bibnamefont{Tian}},
  \bibinfo{author}{\bibfnamefont{X.}~\bibnamefont{Zheng}},
  \bibinfo{author}{\bibfnamefont{T.~J.} \bibnamefont{Kempa}}, \bibnamefont{and}
  \bibinfo{author}{\bibfnamefont{C.~M.} \bibnamefont{Lieber}},
  \bibinfo{journal}{Nature} \textbf{\bibinfo{volume}{449}},
  \bibinfo{pages}{885} (\bibinfo{year}{2007}).

\bibitem[{\citenamefont{Qian et~al.}(2008)\citenamefont{Qian, Li, Grade{\v
  c}ak, Park, Dong, Ding, Wang, and Lieber}}]{Qian:2008cv}
\bibinfo{author}{\bibfnamefont{F.}~\bibnamefont{Qian}},
  \bibinfo{author}{\bibfnamefont{Y.}~\bibnamefont{Li}},
  \bibinfo{author}{\bibfnamefont{S.}~\bibnamefont{Grade{\v c}ak}},
  \bibinfo{author}{\bibfnamefont{H.-G.} \bibnamefont{Park}},
  \bibinfo{author}{\bibfnamefont{Y.}~\bibnamefont{Dong}},
  \bibinfo{author}{\bibfnamefont{Y.}~\bibnamefont{Ding}},
  \bibinfo{author}{\bibfnamefont{Z.~L.} \bibnamefont{Wang}}, \bibnamefont{and}
  \bibinfo{author}{\bibfnamefont{C.~M.} \bibnamefont{Lieber}},
  \bibinfo{journal}{Nature Materials} \textbf{\bibinfo{volume}{7}},
  \bibinfo{pages}{701} (\bibinfo{year}{2008}).

\bibitem[{\citenamefont{Wallentin et~al.}(2013)\citenamefont{Wallentin, Anttu,
  Samuelson, Deppert, and Borgstrom}}]{Wallentin:cp}
\bibinfo{author}{\bibfnamefont{J.}~\bibnamefont{Wallentin}},
  \bibinfo{author}{\bibfnamefont{N.}~\bibnamefont{Anttu}},
  \bibinfo{author}{\bibfnamefont{L.}~\bibnamefont{Samuelson}},
  \bibinfo{author}{\bibfnamefont{K.}~\bibnamefont{Deppert}}, \bibnamefont{and}
  \bibinfo{author}{\bibfnamefont{M.}~\bibnamefont{Borgstrom}},
  \bibinfo{journal}{Science} \textbf{\bibinfo{volume}{339}},
  \bibinfo{pages}{263} (\bibinfo{year}{2013}).

\bibitem[{\citenamefont{Heiss et~al.}(2013)\citenamefont{Heiss, Fontana,
  Gustafsson, W{\"u}st, Magen, O'Regan, Luo, Ketterer, Conesa-Boj, Kuhlmann
  et~al.}}]{Heiss:2013jw}
\bibinfo{author}{\bibfnamefont{M.}~\bibnamefont{Heiss}},
  \bibinfo{author}{\bibfnamefont{Y.}~\bibnamefont{Fontana}},
  \bibinfo{author}{\bibfnamefont{A.}~\bibnamefont{Gustafsson}},
  \bibinfo{author}{\bibfnamefont{G.}~\bibnamefont{W{\"u}st}},
  \bibinfo{author}{\bibfnamefont{C.}~\bibnamefont{Magen}},
  \bibinfo{author}{\bibfnamefont{D.~D.} \bibnamefont{O'Regan}},
  \bibinfo{author}{\bibfnamefont{J.~W.} \bibnamefont{Luo}},
  \bibinfo{author}{\bibfnamefont{B.}~\bibnamefont{Ketterer}},
  \bibinfo{author}{\bibfnamefont{S.}~\bibnamefont{Conesa-Boj}},
  \bibinfo{author}{\bibfnamefont{A.~V.} \bibnamefont{Kuhlmann}},
  \bibnamefont{et~al.}, \bibinfo{journal}{Nature Materials}
  \textbf{\bibinfo{volume}{12}}, \bibinfo{pages}{439} (\bibinfo{year}{2013}).

\bibitem[{\citenamefont{Krogstrup
  et~al.}(2013{\natexlab{a}})\citenamefont{Krogstrup, Jorgensen, Heiss, and
  i~Morral}}]{Krogstrup:mxvdUTp+}
\bibinfo{author}{\bibfnamefont{P.}~\bibnamefont{Krogstrup}},
  \bibinfo{author}{\bibfnamefont{H.}~\bibnamefont{Jorgensen}},
  \bibinfo{author}{\bibfnamefont{M.}~\bibnamefont{Heiss}}, \bibnamefont{and}
  \bibinfo{author}{\bibfnamefont{A.~F.} \bibnamefont{i~Morral}},
  \bibinfo{journal}{Nature Photonics} \textbf{\bibinfo{volume}{7}},
  \bibinfo{pages}{306} (\bibinfo{year}{2013}{\natexlab{a}}).

\bibitem[{\citenamefont{Mourik et~al.}(2012)\citenamefont{Mourik, Zuo, Frolov,
  Plissard, Bakkers, and Kouwenhoven}}]{Mourik:2012je}
\bibinfo{author}{\bibfnamefont{V.}~\bibnamefont{Mourik}},
  \bibinfo{author}{\bibfnamefont{K.}~\bibnamefont{Zuo}},
  \bibinfo{author}{\bibfnamefont{S.}~\bibnamefont{Frolov}},
  \bibinfo{author}{\bibfnamefont{S.}~\bibnamefont{Plissard}},
  \bibinfo{author}{\bibfnamefont{E.}~\bibnamefont{Bakkers}}, \bibnamefont{and}
  \bibinfo{author}{\bibfnamefont{L.}~\bibnamefont{Kouwenhoven}},
  \bibinfo{journal}{Science} \textbf{\bibinfo{volume}{336}},
  \bibinfo{pages}{1003} (\bibinfo{year}{2012}).

\bibitem[{\citenamefont{Deng et~al.}(2012)\citenamefont{Deng, Yu, Huang,
  Larsson, Caroff, and Xu}}]{Deng:2012gn}
\bibinfo{author}{\bibfnamefont{M.}~\bibnamefont{Deng}},
  \bibinfo{author}{\bibfnamefont{C.}~\bibnamefont{Yu}},
  \bibinfo{author}{\bibfnamefont{G.}~\bibnamefont{Huang}},
  \bibinfo{author}{\bibfnamefont{M.}~\bibnamefont{Larsson}},
  \bibinfo{author}{\bibfnamefont{P.}~\bibnamefont{Caroff}}, \bibnamefont{and}
  \bibinfo{author}{\bibfnamefont{H.}~\bibnamefont{Xu}}, \bibinfo{journal}{Nano
  Letters} \textbf{\bibinfo{volume}{12}}, \bibinfo{pages}{6414}
  (\bibinfo{year}{2012}).

\bibitem[{\citenamefont{Schmidt et~al.}(2006)\citenamefont{Schmidt, Riel, Senz,
  and Karg}}]{Schmidt:2006pn}
\bibinfo{author}{\bibfnamefont{V.}~\bibnamefont{Schmidt}},
  \bibinfo{author}{\bibfnamefont{H.}~\bibnamefont{Riel}},
  \bibinfo{author}{\bibfnamefont{S.}~\bibnamefont{Senz}}, \bibnamefont{and}
  \bibinfo{author}{\bibfnamefont{S.}~\bibnamefont{Karg}},
  \bibinfo{journal}{Small} \textbf{\bibinfo{volume}{2}}, \bibinfo{pages}{85}
  (\bibinfo{year}{2006}).

\bibitem[{\citenamefont{Dey et~al.}(2012)\citenamefont{Dey, Thelander, and
  Wernersson}}]{Dey:2012ie}
\bibinfo{author}{\bibfnamefont{A.~W.} \bibnamefont{Dey}},
  \bibinfo{author}{\bibfnamefont{C.}~\bibnamefont{Thelander}},
  \bibnamefont{and}
  \bibinfo{author}{\bibfnamefont{E.}~\bibnamefont{Wernersson}},
  \bibinfo{journal}{IEEE El. Dev. Lett.} \textbf{\bibinfo{volume}{33}},
  \bibinfo{pages}{791} (\bibinfo{year}{2012}).

\bibitem[{\citenamefont{Bessire et~al.}(2011)\citenamefont{Bessire, Bj{\"o}rk,
  Schmid, Schenk, Reuter, and Riel}}]{Bessire:2011gq}
\bibinfo{author}{\bibfnamefont{C.}~\bibnamefont{Bessire}},
  \bibinfo{author}{\bibfnamefont{M.}~\bibnamefont{Bj{\"o}rk}},
  \bibinfo{author}{\bibfnamefont{H.}~\bibnamefont{Schmid}},
  \bibinfo{author}{\bibfnamefont{A.}~\bibnamefont{Schenk}},
  \bibinfo{author}{\bibfnamefont{K.}~\bibnamefont{Reuter}}, \bibnamefont{and}
  \bibinfo{author}{\bibfnamefont{H.}~\bibnamefont{Riel}},
  \bibinfo{journal}{Nano Letters} \textbf{\bibinfo{volume}{11}},
  \bibinfo{pages}{4195} (\bibinfo{year}{2011}).

\bibitem[{\citenamefont{Ohlsson et~al.}(2002)\citenamefont{Ohlsson, Bj{\"o}rk,
  and Persson}}]{Ohlsson:2002qK}
\bibinfo{author}{\bibfnamefont{B.}~\bibnamefont{Ohlsson}},
  \bibinfo{author}{\bibfnamefont{M.}~\bibnamefont{Bj{\"o}rk}},
  \bibnamefont{and} \bibinfo{author}{\bibfnamefont{A.}~\bibnamefont{Persson}},
  \bibinfo{journal}{Physica E} p. \bibinfo{pages}{1126} (\bibinfo{year}{2002}).

\bibitem[{\citenamefont{Dubrovskii and Sibirev}(2008)}]{Dubrovskii:2008tu}
\bibinfo{author}{\bibfnamefont{V.}~\bibnamefont{Dubrovskii}} \bibnamefont{and}
  \bibinfo{author}{\bibfnamefont{N.}~\bibnamefont{Sibirev}},
  \bibinfo{journal}{Phys. Rev. B} \textbf{\bibinfo{volume}{77}},
  \bibinfo{pages}{035414} (\bibinfo{year}{2008}).

\bibitem[{\citenamefont{Uccelli et~al.}(2011)\citenamefont{Uccelli, Arbiol,
  Magen, Krogstrup, Russo-Averchi, Heiss, Mugny, Morier-Genoud, rd, and
  Morante}}]{Uccelli:2011tc}
\bibinfo{author}{\bibfnamefont{E.}~\bibnamefont{Uccelli}},
  \bibinfo{author}{\bibfnamefont{J.}~\bibnamefont{Arbiol}},
  \bibinfo{author}{\bibfnamefont{C.}~\bibnamefont{Magen}},
  \bibinfo{author}{\bibfnamefont{P.}~\bibnamefont{Krogstrup}},
  \bibinfo{author}{\bibfnamefont{E.}~\bibnamefont{Russo-Averchi}},
  \bibinfo{author}{\bibfnamefont{M.}~\bibnamefont{Heiss}},
  \bibinfo{author}{\bibfnamefont{G.}~\bibnamefont{Mugny}},
  \bibinfo{author}{\bibfnamefont{F.}~\bibnamefont{Morier-Genoud}},
  \bibinfo{author}{\bibfnamefont{J.~N.} \bibnamefont{rd}}, \bibnamefont{and}
  \bibinfo{author}{\bibfnamefont{J.}~\bibnamefont{Morante}},
  \bibinfo{journal}{Nano Letters} \textbf{\bibinfo{volume}{11}},
  \bibinfo{pages}{3827} (\bibinfo{year}{2011}).

\bibitem[{\citenamefont{Fang et~al.}(1990)\citenamefont{Fang, Adomi, Iyer,
  Morkoc, Zabel, Choi, and Otsuka}}]{Fang:1990r[}
\bibinfo{author}{\bibfnamefont{S.}~\bibnamefont{Fang}},
  \bibinfo{author}{\bibfnamefont{K.}~\bibnamefont{Adomi}},
  \bibinfo{author}{\bibfnamefont{S.}~\bibnamefont{Iyer}},
  \bibinfo{author}{\bibfnamefont{H.}~\bibnamefont{Morkoc}},
  \bibinfo{author}{\bibfnamefont{H.}~\bibnamefont{Zabel}},
  \bibinfo{author}{\bibfnamefont{C.}~\bibnamefont{Choi}}, \bibnamefont{and}
  \bibinfo{author}{\bibfnamefont{N.}~\bibnamefont{Otsuka}},
  \bibinfo{journal}{Journal of Applied Physics} \textbf{\bibinfo{volume}{68}},
  \bibinfo{pages}{R31} (\bibinfo{year}{1990}).

\bibitem[{\citenamefont{Wagner and Ellis}(1964)}]{Wagner:1964hh}
\bibinfo{author}{\bibfnamefont{R.}~\bibnamefont{Wagner}} \bibnamefont{and}
  \bibinfo{author}{\bibfnamefont{W.}~\bibnamefont{Ellis}},
  \bibinfo{journal}{Applied Physics Letters} \textbf{\bibinfo{volume}{4}},
  \bibinfo{pages}{89} (\bibinfo{year}{1964}).

\bibitem[{\citenamefont{Dubrovskii et~al.}(2009)\citenamefont{Dubrovskii,
  Cirlin, and Ustinov}}]{Dubrovskii:2009sm}
\bibinfo{author}{\bibfnamefont{V.}~\bibnamefont{Dubrovskii}},
  \bibinfo{author}{\bibfnamefont{G.}~\bibnamefont{Cirlin}}, \bibnamefont{and}
  \bibinfo{author}{\bibfnamefont{V.}~\bibnamefont{Ustinov}},
  \bibinfo{journal}{Semiconductors} \textbf{\bibinfo{volume}{43}},
  \bibinfo{pages}{1539} (\bibinfo{year}{2009}).

\bibitem[{\citenamefont{Colombo et~al.}(2008)\citenamefont{Colombo, Spirkoska,
  Frimmer, Abstreiter, and i~Morral}}]{Colombo:2008ci}
\bibinfo{author}{\bibfnamefont{C.}~\bibnamefont{Colombo}},
  \bibinfo{author}{\bibfnamefont{D.}~\bibnamefont{Spirkoska}},
  \bibinfo{author}{\bibfnamefont{M.}~\bibnamefont{Frimmer}},
  \bibinfo{author}{\bibfnamefont{G.}~\bibnamefont{Abstreiter}},
  \bibnamefont{and} \bibinfo{author}{\bibfnamefont{A.~F.}
  \bibnamefont{i~Morral}}, \bibinfo{journal}{Phys. Rev. B}
  \textbf{\bibinfo{volume}{77}}, \bibinfo{pages}{155326}
  (\bibinfo{year}{2008}).

\bibitem[{\citenamefont{Paek et~al.}(2009)\citenamefont{Paek, Nishiwaki,
  Yamaguchi, and Sawaki}}]{Paek:2009kr}
\bibinfo{author}{\bibfnamefont{J.~H.} \bibnamefont{Paek}},
  \bibinfo{author}{\bibfnamefont{T.}~\bibnamefont{Nishiwaki}},
  \bibinfo{author}{\bibfnamefont{M.}~\bibnamefont{Yamaguchi}},
  \bibnamefont{and} \bibinfo{author}{\bibfnamefont{N.}~\bibnamefont{Sawaki}},
  \bibinfo{journal}{Phys. Stat. Sol. (c)} \textbf{\bibinfo{volume}{6}},
  \bibinfo{pages}{1436} (\bibinfo{year}{2009}).

\bibitem[{\citenamefont{Sadowski et~al.}(2008)\citenamefont{Sadowski,
  Dluzewski, and Kanski}}]{Sadowski:JSQlMZtc}
\bibinfo{author}{\bibfnamefont{J.}~\bibnamefont{Sadowski}},
  \bibinfo{author}{\bibfnamefont{P.}~\bibnamefont{Dluzewski}},
  \bibnamefont{and} \bibinfo{author}{\bibfnamefont{J.}~\bibnamefont{Kanski}},
  p. \bibinfo{pages}{arXiv:0812.2453} (\bibinfo{year}{2008}).

\bibitem[{\citenamefont{Heiss et~al.}(2008)\citenamefont{Heiss, Riedlberger,
  Spirkoska, Bichler, Abstrelter, and i~Morral}}]{Heiss:2008ff}
\bibinfo{author}{\bibfnamefont{M.}~\bibnamefont{Heiss}},
  \bibinfo{author}{\bibfnamefont{E.}~\bibnamefont{Riedlberger}},
  \bibinfo{author}{\bibfnamefont{D.}~\bibnamefont{Spirkoska}},
  \bibinfo{author}{\bibfnamefont{M.}~\bibnamefont{Bichler}},
  \bibinfo{author}{\bibfnamefont{G.}~\bibnamefont{Abstrelter}},
  \bibnamefont{and} \bibinfo{author}{\bibfnamefont{A.~F.}
  \bibnamefont{i~Morral}}, \bibinfo{journal}{J Cryst Growth}
  \textbf{\bibinfo{volume}{310}}, \bibinfo{pages}{1049} (\bibinfo{year}{2008}).

\bibitem[{\citenamefont{Jabeen et~al.}(2008)\citenamefont{Jabeen, Grillo,
  Rubini, and Martelli}}]{Jabeen:2008tZ}
\bibinfo{author}{\bibfnamefont{F.}~\bibnamefont{Jabeen}},
  \bibinfo{author}{\bibfnamefont{V.}~\bibnamefont{Grillo}},
  \bibinfo{author}{\bibfnamefont{S.}~\bibnamefont{Rubini}}, \bibnamefont{and}
  \bibinfo{author}{\bibfnamefont{F.}~\bibnamefont{Martelli}},
  \bibinfo{journal}{Nanotechnology} \textbf{\bibinfo{volume}{19}},
  \bibinfo{pages}{275711} (\bibinfo{year}{2008}).

\bibitem[{\citenamefont{Breuer et~al.}(2011)\citenamefont{Breuer, Pfuller,
  Flissikowski, Brandt, Grahn, Geelhaar, and Riechert}}]{Breuer:2011gx}
\bibinfo{author}{\bibfnamefont{S.}~\bibnamefont{Breuer}},
  \bibinfo{author}{\bibfnamefont{C.}~\bibnamefont{Pfuller}},
  \bibinfo{author}{\bibfnamefont{T.}~\bibnamefont{Flissikowski}},
  \bibinfo{author}{\bibfnamefont{O.}~\bibnamefont{Brandt}},
  \bibinfo{author}{\bibfnamefont{H.}~\bibnamefont{Grahn}},
  \bibinfo{author}{\bibfnamefont{L.}~\bibnamefont{Geelhaar}}, \bibnamefont{and}
  \bibinfo{author}{\bibfnamefont{H.}~\bibnamefont{Riechert}},
  \bibinfo{journal}{Nano Letters} \textbf{\bibinfo{volume}{11}},
  \bibinfo{pages}{1276} (\bibinfo{year}{2011}).

\bibitem[{\citenamefont{Ramdani et~al.}(2013)\citenamefont{Ramdani, Harmand,
  Glas, Patriarche, and Travers}}]{Anonymous:2012gc}
\bibinfo{author}{\bibfnamefont{M.~R.} \bibnamefont{Ramdani}},
  \bibinfo{author}{\bibfnamefont{J.~C.} \bibnamefont{Harmand}},
  \bibinfo{author}{\bibfnamefont{F.}~\bibnamefont{Glas}},
  \bibinfo{author}{\bibfnamefont{G.}~\bibnamefont{Patriarche}},
  \bibnamefont{and} \bibinfo{author}{\bibfnamefont{L.}~\bibnamefont{Travers}},
  \bibinfo{journal}{Crystal Growth {\&} Design} \textbf{\bibinfo{volume}{13}},
  \bibinfo{pages}{91} (\bibinfo{year}{2013}).

\bibitem[{\citenamefont{Krogstrup
  et~al.}(2013{\natexlab{b}})\citenamefont{Krogstrup, Jorgensen, i~Morral, and
  Glas}}]{Krogstrup:2013tl}
\bibinfo{author}{\bibfnamefont{P.}~\bibnamefont{Krogstrup}},
  \bibinfo{author}{\bibnamefont{Jorgensen}},
  \bibinfo{author}{\bibfnamefont{A.~F.} \bibnamefont{i~Morral}},
  \bibnamefont{and} \bibinfo{author}{\bibfnamefont{F.}~\bibnamefont{Glas}},
  \bibinfo{journal}{J. Phys. D: Appl. Phys.} \textbf{\bibinfo{volume}{46}},
  \bibinfo{pages}{313001} (\bibinfo{year}{2013}{\natexlab{b}}).

\bibitem[{\citenamefont{Krogstrup et~al.}(2010)\citenamefont{Krogstrup,
  Popovitz-Biro, Johnson, and Shtrikman}}]{Krogstrup:2010ux}
\bibinfo{author}{\bibfnamefont{P.}~\bibnamefont{Krogstrup}},
  \bibinfo{author}{\bibfnamefont{R.}~\bibnamefont{Popovitz-Biro}},
  \bibinfo{author}{\bibfnamefont{E.}~\bibnamefont{Johnson}}, \bibnamefont{and}
  \bibinfo{author}{\bibfnamefont{H.}~\bibnamefont{Shtrikman}},
  \bibinfo{journal}{Nano Letters} \textbf{\bibinfo{volume}{10}},
  \bibinfo{pages}{4475} (\bibinfo{year}{2010}).

\bibitem[{\citenamefont{Plissard et~al.}(2010)\citenamefont{Plissard, Dick,
  Larrieu, and Godey}}]{Plissard:2010qO}
\bibinfo{author}{\bibfnamefont{S.}~\bibnamefont{Plissard}},
  \bibinfo{author}{\bibfnamefont{K.}~\bibnamefont{Dick}},
  \bibinfo{author}{\bibfnamefont{G.}~\bibnamefont{Larrieu}}, \bibnamefont{and}
  \bibinfo{author}{\bibfnamefont{S.}~\bibnamefont{Godey}},
  \bibinfo{journal}{Nanotechnology} \textbf{\bibinfo{volume}{21}},
  \bibinfo{pages}{385602} (\bibinfo{year}{2010}).

\bibitem[{\citenamefont{Plissard et~al.}(2011)\citenamefont{Plissard, Larrieu,
  Wallart, and Caroff}}]{Plissard:YNqQ9GDR}
\bibinfo{author}{\bibfnamefont{S.}~\bibnamefont{Plissard}},
  \bibinfo{author}{\bibfnamefont{G.}~\bibnamefont{Larrieu}},
  \bibinfo{author}{\bibfnamefont{X.}~\bibnamefont{Wallart}}, \bibnamefont{and}
  \bibinfo{author}{\bibfnamefont{P.}~\bibnamefont{Caroff}},
  \bibinfo{journal}{Nanotechnology} \textbf{\bibinfo{volume}{22}},
  \bibinfo{pages}{275602} (\bibinfo{year}{2011}).

\bibitem[{\citenamefont{Samsonenko et~al.}(2011)\citenamefont{Samsonenko,
  Cirlin, Khrebtov, and Werner}}]{Samsonenko:2011rN}
\bibinfo{author}{\bibfnamefont{Y.}~\bibnamefont{Samsonenko}},
  \bibinfo{author}{\bibfnamefont{G.}~\bibnamefont{Cirlin}},
  \bibinfo{author}{\bibfnamefont{A.}~\bibnamefont{Khrebtov}}, \bibnamefont{and}
  \bibinfo{author}{\bibfnamefont{P.}~\bibnamefont{Werner}},
  \bibinfo{journal}{Semiconductors} \textbf{\bibinfo{volume}{45}},
  \bibinfo{pages}{431} (\bibinfo{year}{2011}).

\bibitem[{\citenamefont{Neuwald et~al.}(1992)\citenamefont{Neuwald, Hessel,
  Feltz, Memmert, and Behm}}]{Neuwald:sk2aXoay}
\bibinfo{author}{\bibfnamefont{U.}~\bibnamefont{Neuwald}},
  \bibinfo{author}{\bibfnamefont{H.~E.} \bibnamefont{Hessel}},
  \bibinfo{author}{\bibfnamefont{A.}~\bibnamefont{Feltz}},
  \bibinfo{author}{\bibfnamefont{U.}~\bibnamefont{Memmert}}, \bibnamefont{and}
  \bibinfo{author}{\bibfnamefont{R.~J.} \bibnamefont{Behm}},
  \bibinfo{journal}{Applied Physics Letters} \textbf{\bibinfo{volume}{60}},
  \bibinfo{pages}{1} (\bibinfo{year}{1992}).

\bibitem[{\citenamefont{Burrows et~al.}(1988)\citenamefont{Burrows, Chabal,
  Higashi, Raghavachari, and Christman}}]{Burrows:1988hl}
\bibinfo{author}{\bibfnamefont{A.}~\bibnamefont{Burrows}},
  \bibinfo{author}{\bibfnamefont{Y.}~\bibnamefont{Chabal}},
  \bibinfo{author}{\bibfnamefont{G.}~\bibnamefont{Higashi}},
  \bibinfo{author}{\bibfnamefont{K.}~\bibnamefont{Raghavachari}},
  \bibnamefont{and}
  \bibinfo{author}{\bibfnamefont{S.}~\bibnamefont{Christman}},
  \bibinfo{journal}{Applied Physics Letters} \textbf{\bibinfo{volume}{53}},
  \bibinfo{pages}{998} (\bibinfo{year}{1988}).

\bibitem[{\citenamefont{Soria et~al.}(2012)\citenamefont{Soria, Patrito, and
  Olivera}}]{Soria:2012gm}
\bibinfo{author}{\bibfnamefont{F.~A.} \bibnamefont{Soria}},
  \bibinfo{author}{\bibfnamefont{E.~M.} \bibnamefont{Patrito}},
  \bibnamefont{and} \bibinfo{author}{\bibfnamefont{P.~P.}
  \bibnamefont{Olivera}}, \bibinfo{journal}{The Journal of Physical Chemistry
  C} \textbf{\bibinfo{volume}{116}}, \bibinfo{pages}{24607}
  (\bibinfo{year}{2012}).

\bibitem[{\citenamefont{i~Morral et~al.}(2008)\citenamefont{i~Morral, Colombo,
  and Morante}}]{Anonymous:jlM9QbYX}
\bibinfo{author}{\bibfnamefont{A.~F.} \bibnamefont{i~Morral}},
  \bibinfo{author}{\bibfnamefont{C.}~\bibnamefont{Colombo}}, \bibnamefont{and}
  \bibinfo{author}{\bibfnamefont{J.}~\bibnamefont{Morante}},
  \bibinfo{journal}{Applied Physics Letters} \textbf{\bibinfo{volume}{92}},
  \bibinfo{pages}{063112} (\bibinfo{year}{2008}).

\bibitem[{\citenamefont{Stangl et~al.}(2010)\citenamefont{Stangl, Mandl,
  Hilner, Zakharov, Hillerich, Dey, Samuelson, Deppert, and
  Mikkelsen}}]{Stangl:2010bh}
\bibinfo{author}{\bibfnamefont{J.}~\bibnamefont{Stangl}},
  \bibinfo{author}{\bibfnamefont{B.}~\bibnamefont{Mandl}},
  \bibinfo{author}{\bibfnamefont{E.}~\bibnamefont{Hilner}},
  \bibinfo{author}{\bibfnamefont{A.~A.} \bibnamefont{Zakharov}},
  \bibinfo{author}{\bibfnamefont{K.}~\bibnamefont{Hillerich}},
  \bibinfo{author}{\bibfnamefont{A.~W.} \bibnamefont{Dey}},
  \bibinfo{author}{\bibfnamefont{L.}~\bibnamefont{Samuelson}},
  \bibinfo{author}{\bibfnamefont{K.}~\bibnamefont{Deppert}}, \bibnamefont{and}
  \bibinfo{author}{\bibfnamefont{A.}~\bibnamefont{Mikkelsen}},
  \bibinfo{journal}{Nano Letters} \textbf{\bibinfo{volume}{10}},
  \bibinfo{pages}{1} (\bibinfo{year}{2010}).

\bibitem[{\citenamefont{Boyd}(1982)}]{Boyd:1982kn}
\bibinfo{author}{\bibfnamefont{I.}~\bibnamefont{Boyd}},
  \bibinfo{journal}{Journal of Applied Physics} \textbf{\bibinfo{volume}{53}},
  \bibinfo{pages}{4166} (\bibinfo{year}{1982}).

\bibitem[{\citenamefont{Pai}(1986)}]{Pai:fz}
\bibinfo{author}{\bibfnamefont{P.}~\bibnamefont{Pai}},
  \bibinfo{journal}{Journal of Vacuum Science {\&} Technology A: Vacuum,
  Surfaces, and Films} \textbf{\bibinfo{volume}{4}}, \bibinfo{pages}{689}
  (\bibinfo{year}{1986}).

\bibitem[{\citenamefont{Kim et~al.}(2003)\citenamefont{Kim, Ahn, and
  Ahn}}]{Kim:2003br}
\bibinfo{author}{\bibfnamefont{B.}~\bibnamefont{Kim}},
  \bibinfo{author}{\bibfnamefont{J.}~\bibnamefont{Ahn}}, \bibnamefont{and}
  \bibinfo{author}{\bibfnamefont{B.}~\bibnamefont{Ahn}},
  \bibinfo{journal}{Applied Physics Letters} \textbf{\bibinfo{volume}{82}},
  \bibinfo{pages}{2682} (\bibinfo{year}{2003}).

\bibitem[{\citenamefont{Ono et~al.}(1998)\citenamefont{Ono, Ikarashi, Ando, and
  Kitano}}]{Ono:1998jf}
\bibinfo{author}{\bibfnamefont{H.}~\bibnamefont{Ono}},
  \bibinfo{author}{\bibfnamefont{T.}~\bibnamefont{Ikarashi}},
  \bibinfo{author}{\bibfnamefont{K.}~\bibnamefont{Ando}}, \bibnamefont{and}
  \bibinfo{author}{\bibfnamefont{T.}~\bibnamefont{Kitano}},
  \bibinfo{journal}{Journal of Applied Physics} \textbf{\bibinfo{volume}{84}},
  \bibinfo{pages}{6064} (\bibinfo{year}{1998}).

\bibitem[{\citenamefont{Tian et~al.}(2010)\citenamefont{Tian, Seitz, Li, Hu,
  Chabal, and Gao}}]{Tian:2010iz}
\bibinfo{author}{\bibfnamefont{R.}~\bibnamefont{Tian}},
  \bibinfo{author}{\bibfnamefont{O.}~\bibnamefont{Seitz}},
  \bibinfo{author}{\bibfnamefont{M.}~\bibnamefont{Li}},
  \bibinfo{author}{\bibfnamefont{W.}~\bibnamefont{Hu}},
  \bibinfo{author}{\bibfnamefont{Y.}~\bibnamefont{Chabal}}, \bibnamefont{and}
  \bibinfo{author}{\bibfnamefont{J.}~\bibnamefont{Gao}},
  \bibinfo{journal}{Langmuir Letter} \textbf{\bibinfo{volume}{26}},
  \bibinfo{pages}{4563} (\bibinfo{year}{2010}).

\bibitem[{\citenamefont{Weldon et~al.}(1999)\citenamefont{Weldon, Queeney,
  Chabal, Stefanov, and Raghavachari}}]{Weldon:1999hf}
\bibinfo{author}{\bibfnamefont{M.}~\bibnamefont{Weldon}},
  \bibinfo{author}{\bibfnamefont{K.}~\bibnamefont{Queeney}},
  \bibinfo{author}{\bibfnamefont{Y.}~\bibnamefont{Chabal}},
  \bibinfo{author}{\bibfnamefont{B.}~\bibnamefont{Stefanov}}, \bibnamefont{and}
  \bibinfo{author}{\bibfnamefont{K.}~\bibnamefont{Raghavachari}},
  \bibinfo{journal}{Journal of Vacuum Science {\&} Technology B:
  Microelectronics and Nanometer Structures} \textbf{\bibinfo{volume}{17}},
  \bibinfo{pages}{1795} (\bibinfo{year}{1999}).

\bibitem[{\citenamefont{Queeney et~al.}(2000)\citenamefont{Queeney, Weldon,
  Chang, Chabal, Gurevich, Sapjeta, and Opila}}]{Queeney:2000km}
\bibinfo{author}{\bibfnamefont{K.~T.} \bibnamefont{Queeney}},
  \bibinfo{author}{\bibfnamefont{M.~K.} \bibnamefont{Weldon}},
  \bibinfo{author}{\bibfnamefont{J.~P.} \bibnamefont{Chang}},
  \bibinfo{author}{\bibfnamefont{Y.~J.} \bibnamefont{Chabal}},
  \bibinfo{author}{\bibfnamefont{A.~B.} \bibnamefont{Gurevich}},
  \bibinfo{author}{\bibfnamefont{J.}~\bibnamefont{Sapjeta}}, \bibnamefont{and}
  \bibinfo{author}{\bibfnamefont{R.~L.} \bibnamefont{Opila}},
  \bibinfo{journal}{Journal of Applied Physics} \textbf{\bibinfo{volume}{87}},
  \bibinfo{pages}{1322} (\bibinfo{year}{2000}).

\bibitem[{\citenamefont{Ogawa et~al.}(1996)\citenamefont{Ogawa, Ishikawa, and
  Inomata}}]{Anonymous:cDm1LXIV}
\bibinfo{author}{\bibfnamefont{H.}~\bibnamefont{Ogawa}},
  \bibinfo{author}{\bibfnamefont{K.}~\bibnamefont{Ishikawa}}, \bibnamefont{and}
  \bibinfo{author}{\bibfnamefont{C.}~\bibnamefont{Inomata}},
  \bibinfo{journal}{Journal of Applied Physics} \textbf{\bibinfo{volume}{79}},
  \bibinfo{pages}{472} (\bibinfo{year}{1996}).

\bibitem[{\citenamefont{Ohmi et~al.}(1992)\citenamefont{Ohmi, Morita, Teramoto,
  Makihara, and Tseng}}]{Ohmi:ihawLB9H}
\bibinfo{author}{\bibfnamefont{T.}~\bibnamefont{Ohmi}},
  \bibinfo{author}{\bibfnamefont{M.}~\bibnamefont{Morita}},
  \bibinfo{author}{\bibfnamefont{A.}~\bibnamefont{Teramoto}},
  \bibinfo{author}{\bibfnamefont{K.}~\bibnamefont{Makihara}}, \bibnamefont{and}
  \bibinfo{author}{\bibfnamefont{K.~S.} \bibnamefont{Tseng}},
  \bibinfo{journal}{Applied Physics Letters} \textbf{\bibinfo{volume}{60}},
  \bibinfo{pages}{1} (\bibinfo{year}{1992}).

\bibitem[{\citenamefont{Jakob et~al.}(1991)\citenamefont{Jakob, Dumas, and
  Chabal}}]{Anonymous:qSDNqIRs}
\bibinfo{author}{\bibfnamefont{P.}~\bibnamefont{Jakob}},
  \bibinfo{author}{\bibfnamefont{P.}~\bibnamefont{Dumas}}, \bibnamefont{and}
  \bibinfo{author}{\bibfnamefont{Y.}~\bibnamefont{Chabal}},
  \bibinfo{journal}{Applied Physics Letters} \textbf{\bibinfo{volume}{59}},
  \bibinfo{pages}{2968} (\bibinfo{year}{1991}).

\bibitem[{\citenamefont{Muller et~al.}(1999)\citenamefont{Muller, Sorsch,
  Moccio, and Baumann}}]{Muller:4ei7Orkm}
\bibinfo{author}{\bibnamefont{Muller}},
  \bibinfo{author}{\bibfnamefont{T.}~\bibnamefont{Sorsch}},
  \bibinfo{author}{\bibfnamefont{S.}~\bibnamefont{Moccio}}, \bibnamefont{and}
  \bibinfo{author}{\bibfnamefont{F.}~\bibnamefont{Baumann}},
  \bibinfo{journal}{Nature} \textbf{\bibinfo{volume}{399}},
  \bibinfo{pages}{758} (\bibinfo{year}{1999}).

\bibitem[{\citenamefont{Morita et~al.}(1990)\citenamefont{Morita, Ohmi,
  Hasegawa, and Ohwada}}]{Anonymous:BSAUS+Pe}
\bibinfo{author}{\bibfnamefont{M.}~\bibnamefont{Morita}},
  \bibinfo{author}{\bibfnamefont{T.}~\bibnamefont{Ohmi}},
  \bibinfo{author}{\bibfnamefont{E.}~\bibnamefont{Hasegawa}}, \bibnamefont{and}
  \bibinfo{author}{\bibfnamefont{M.}~\bibnamefont{Ohwada}},
  \bibinfo{journal}{Journal of Applied Physics} \textbf{\bibinfo{volume}{68}},
  \bibinfo{pages}{1272} (\bibinfo{year}{1990}).

\bibitem[{\citenamefont{Al-Bayati et~al.}(1991)\citenamefont{Al-Bayati,
  Orrman-Rossiter, and Berg}}]{Anonymous:1kXBocAk}
\bibinfo{author}{\bibfnamefont{A.}~\bibnamefont{Al-Bayati}},
  \bibinfo{author}{\bibfnamefont{K.}~\bibnamefont{Orrman-Rossiter}},
  \bibnamefont{and} \bibinfo{author}{\bibfnamefont{J.}~\bibnamefont{Berg}},
  \bibinfo{journal}{Surface Science} \textbf{\bibinfo{volume}{241}},
  \bibinfo{pages}{91} (\bibinfo{year}{1991}).

\bibitem[{\citenamefont{Albrecht}(1998)}]{Albrecht:1998iu}
\bibinfo{author}{\bibfnamefont{M.}~\bibnamefont{Albrecht}},
  \bibinfo{journal}{J. Electrochem. Soc.} \textbf{\bibinfo{volume}{145}},
  \bibinfo{pages}{4019} (\bibinfo{year}{1998}).

\bibitem[{\citenamefont{Yang and Chen}(2002)}]{Yang:91OsoeUQ}
\bibinfo{author}{\bibfnamefont{C.}~\bibnamefont{Yang}} \bibnamefont{and}
  \bibinfo{author}{\bibfnamefont{W.}~\bibnamefont{Chen}}, \bibinfo{journal}{J.
  Mater. Chem.} \textbf{\bibinfo{volume}{12}}, \bibinfo{pages}{1138}
  (\bibinfo{year}{2002}).

\bibitem[{\citenamefont{Giang et~al.}(2013)\citenamefont{Giang, Bougerol,
  Mariette, and Songmuang}}]{Giang:2012tx}
\bibinfo{author}{\bibnamefont{Giang}},
  \bibinfo{author}{\bibnamefont{Bougerol}},
  \bibinfo{author}{\bibnamefont{Mariette}}, \bibnamefont{and}
  \bibinfo{author}{\bibnamefont{Songmuang}}, \bibinfo{journal}{Journal of
  Crystal Growth} \textbf{\bibinfo{volume}{364}}, \bibinfo{pages}{118}
  (\bibinfo{year}{2013}).

\bibitem[{\citenamefont{Russo-Averchi et~al.}(2012)\citenamefont{Russo-Averchi,
  Heiss, Michelet, and Krogstrup}}]{RussoAverchi:2012rZ}
\bibinfo{author}{\bibfnamefont{E.}~\bibnamefont{Russo-Averchi}},
  \bibinfo{author}{\bibfnamefont{M.}~\bibnamefont{Heiss}},
  \bibinfo{author}{\bibfnamefont{L.}~\bibnamefont{Michelet}}, \bibnamefont{and}
  \bibinfo{author}{\bibfnamefont{P.}~\bibnamefont{Krogstrup}},
  \bibinfo{journal}{Nanoscale} \textbf{\bibinfo{volume}{4}},
  \bibinfo{pages}{1486} (\bibinfo{year}{2012}).

\bibitem[{\citenamefont{Rieger et~al.}(2012)\citenamefont{Rieger, Heiderich,
  and Lenk}}]{Rieger:UY13hl5a}
\bibinfo{author}{\bibfnamefont{T.}~\bibnamefont{Rieger}},
  \bibinfo{author}{\bibfnamefont{S.}~\bibnamefont{Heiderich}},
  \bibnamefont{and} \bibinfo{author}{\bibfnamefont{S.}~\bibnamefont{Lenk}},
  \bibinfo{journal}{Journal of Crystal Growth} \textbf{\bibinfo{volume}{353}},
  \bibinfo{pages}{39} (\bibinfo{year}{2012}).

\end{thebibliography}

\end{document}